\documentclass[5p]{elsarticle}
\biboptions{sort&compress}

\usepackage{amssymb}
\usepackage{amsthm}

\usepackage{graphicx}%
\usepackage{dcolumn}%
\usepackage{bm}%
\usepackage{hyperref}%
\newcommand{\Jsc}{{$J_\mathrm{sc}$~}}
\newcommand{\Voc}{{$V_\mathrm{oc}$~}}

\makeatletter
\def\ps@pprintTitle{%
  \let\@oddhead\@empty
  \let\@evenhead\@empty
  \def\@oddfoot{\reset@font\hfil\thepage\hfil}
  \let\@evenfoot\@oddfoot
}
\makeatother

\begin{document}
\title{Efficient light-trapping in ultrathin GaAs solar cells using quasi-random photonic crystals}

\author[nrel,ipvf,c2n]{Jeronimo Buencuerpo\corref{cor1}}
\ead{jeronimo.buencuerpo@ipvf.fr}
\author[nrel]{Theresa E. Saenz} 
\author[nrel]{Mark Steger} 
\author[nrel]{Michelle Young}
\author[nrel]{Emily L. Warren}
\author[nrel]{John F. Geisz}
\author[nrel]{Myles A. Steiner}
\author[nrel]{Adele C. Tamboli}
\cortext[cor1]{Corresponding author}

\address[nrel]{National Renewable Energy Laboratory (NREL), Golden 80401 Colorado, USA}
\address[ipvf]{Present address at L'Institut Photovoltaïque d'Île-de-France (IPVF), Palaiseau, France}
\address[c2n]{Present address at Centre de Nanosciences and Nanotechnologies (C2N), CNRS, Paris-Saclay University, Palaiseau, France}

\begin{abstract}
Ultrathin solar cells reduce material usage and allow the use of lower-quality materials thanks to their one order of magnitude smaller thickness than their conventional counterparts. However, efficient photonic light-trapping is required to harvest the incident light efficiently for an otherwise insufficient absorber thickness. Quasi-random photonic crystals are predicted to have high efficient light-trapping while being more robust under angle and thickness variations than simple photonic crystals. Here we experimentally demonstrate a light-trapping solution based on quasi-random photonic crystals fabricated by polymer blend lithography. We control the average lattice parameter by modifying the spin-coating speed. We demonstrate an ultrathin GaAs cell of 260 nm with a rear quasi-random pattern with submicron features, and a \Jsc=26.4 mA/cm$^2$ and an efficiency of 22.35\% under the global solar spectrum.
\end{abstract}
\maketitle

\begin{keyword}
photovoltaics, quasi-random, photonic crystals, ultrathin, GaAs
\end{keyword}

\section{Introduction}
The potential of nanotechnology to further push the limits of current photovoltaic technology has been repeatedly demonstrated in the past two decades\cite{garnett_2021_ap,polman_2012_nm,cariou_2018_ne,steiner_2021_aem,brongersma_2014_nm}. The use of photonic crystals for light trapping has been a recurrent theme in these studies, but such advances have so far failed to materialize as commercial products, largely due to the need for capital intensive fabrication methods such as nanolithography and dry etching. Here we demonstrate that quasi-random photonic structures (apparently disordered structures with a precisely defined reciprocal space distribution of spatial frequencies) lead to a 10\% relative improvement of ultrathin solar cells, allowing for a significant reduction in the material cost of photovoltaic devices without using expensive fabrication techniques.

Ultrathin solar cells typically have an active layer that is an order of magnitude (or more) thinner than their thick counterparts \cite{massiot_2020_ne}. The thinner active layer presents the immediate benefit of lowering the material usage, potentially improving the economics of the solar cell. 
This will benefit cells made with more expensive materials, like the ones based on III-Vs, or minimize the use of rare or toxic materials. Additionally, one of the benefits of ultrathin cells is that they relax the quality of the material needed for obtaining a high-efficiency cell. Namely, the diffusion length can be an order of magnitude less than in thick cells while maintaining similar efficiency. This will directly benefit, for example, a strained growth or a polycrystalline cell\cite{mehrotra_2014_2i4pscp,greenaway_2017_ael}, and will improve the radiation hardness of space cells operating in harsh radiation conditions\cite{yamaguchi_2001_semasc,hirst_2016_apl,maximenko_2019_2i4pscp}. In space applications, the specific power (W/kg) is an additional fundamental figure of metric. Ultrathin cells can outperform their thicker counterparts because, if radiation hardness is achieved, the thickness of coverglass used for protection can be thinned down, thus lowering the final weight of the module.
However, ultrathin cells require optimized light-trapping structures to achieve comparable photocurrents and efficiencies to their thicker counterparts\cite{chen_2019_ne,eerden_2020_ppra,sayre_2022_ppra}.  Therefore, developing high-efficiency and cost-effective light trapping methods is fundamental for ultrathin cells. 

Light-trapping can be achieved by using photonic crystals, repeating a unit cell in a plane made of two or more materials with different permittivities. The repetition of the unit cell has as a consequence a reciprocal vector associated with the lattice that adds momentum to the incident light, therefore changing the optical path\cite{tikhodeev_2002_prb}. These structures are highly diffractive, but typically require an optimal tuning of the size parameters as the resonances are sharp. On the other hand, there is the possibility of using completely rough structures, without a defined lattice. These patterns must have scatterers of comparable size to the incident light, and each one is able to re-direct the incident light, increasing its optical path. While these structures have more broadband resonances, they also exhibit a lower spectral density\cite{vanlare_2015_ap}. The ideal structure should compromise the higher power spectral density (PSD) of simple photonic crystals, and the broader operating wavelengths of the random structures.
Quasi-random (also known as hyperuniform) photonic crystals (QR-PCs) are structures with a reciprocal space defined by a high diffraction efficiency with power spectral density localized over a ring\cite{ma_2017_joap,martins_2013_nca,yu_2017_sr,buencuerpo_2021_olt}, but zero or nearly-zero diffraction efficiency inside and outside the ring. The higher PSD implies a higher diffraction efficiency similar to the simple photonic crystals. Also, the ring PSD has the advantage of accommodating a broader number of spectral frequencies than simple photonic crystal such as squares or cylinders in square/hexagonal lattices\cite{vanlare_2015_ap,abad_2020_oeo,buencuerpo_2021_olt}, and as a consequence, accommodates a wider operating wavelength range than simple photonic crystals\cite{buencuerpo_2021_olt}. The reciprocal space defines these structures, and therefore the real space can be formed with different geometries, such as intertwined structures\cite{buencuerpo_2021_olt,vanlare_2015_ap,martins_2013_nca}, or individual structures organized like pillars or ovals\cite{yu_2017_sr,gorsky_2019_ap,castro-lopez_2017_ap}.

For ultrathin GaAs there are two main approaches: using simple periodic structures\cite{chen_2019_ne}, or roughening by using anisotropic wet etching\cite{eerden_2020_ppra,drozario_2020_ijp}. The simple periodic approach has the advantage of obtaining the optimal scatterers if the design is accurate, while using a pre-produced stamp with a higher cost. On the other hand, the anisotropic wet etching has the advantage of being mask-less and cheaper but the scatterers are not optimal, limiting the photocurrent improvements. Quasi-random photonic crystals (QR-PCs) have been proposed on passive substrates and fabricated using electron beam lithography\cite{li_2016_sr,xiao_2018_oe}, and using wrinkle lithography combined with reactive ion etching\cite{lee_2017_p}. Another option is to use polymer blends (PBs) to create the quasi-random pattern\cite{swalheim_1999_s,huang_2012_bjn,guo_2015_sr,zhang_2018_oeo,donie_2021_an}. The pattern is created by mixing two immiscible polymers in a common solvent. When the solvent evaporates, a spinodal distribution following the Can-Hillard equation\cite{henderson_2004_m} generates the average lattice of the structure with domains defined by the ratio of polymer volumes. The lattice is typically hundreds of nanometers to tens of microns\cite{swalheim_1999_s,zhang_2018_oeo}, but it can be smaller as has been shown for block copolymer lithography\cite{kim_2010_cr}. The reciprocal lattice of the photonic crystal should have sizes close to the GaAs frequency bandgap to effectively diffract the useful wavelengths for the GaAs solar cell, or a few multiples of it (when exciting diffraction orders above the first); therefore, we are aiming for a lattice between 600-900 nm\cite{chen_2019_ne, buencuerpo_2020_oe}. Short-wavelengths (below 500 nm for an ultrathin GaAs cell) do not require scattering as the cell is optically thick at those wavelengths.%

Polymer blend lithography has been proposed for patterned contacts in solar cells, burying one of the polymers\cite{donie_2018_n,hauser_2020_oeo}, or combined with reactive ion etching (RIE) to transfer the pattern to the semiconductor\cite{siddique_2017_sa,zhang_2018_oeo,guo_2015_sr}. Covering the polymer with other materials ties the lattice obtained by spin-coating to the polymer's coating thickness (typically under 100 nm), leaving little room for tuning the structures and limiting its diffraction properties. Using RIE can alleviate this limitation; when using a selective etch, the structure thickness is not tied to the lattice from the coating. However, the use of RIE in photovoltaics is not as extensive as in other areas of electronics, limiting its impact. Here, we combine PB lithography with simple wet etching, using the polymer as a mask, to transfer the pattern into the semiconductor. This is possible due to the vertical orientation of the polymer domains and the high wet-etching selectivity between the polymer and the semiconductor.

The paper is organized as follows. In Section \ref{sec:design} we describe the overall design choices made for an ultrathin GaAs solar cell, including modifications to the baseline inverted growth. In Section \ref{sec:fab_qr} we describe the fabrication process using a polymer blend to obtain random microstructures and two quasi-random photonic crystals (QR-PCs) in AlGaAs. In section \ref{sec:fab_cell} we describe the experimental study of ultrathin GaAs cells for a planar cell, a random microstructure, and two QR-PCs. %
Finally, we discuss the results by analyzing the origin of the enhancement in photocurrent in these ultrathin devices.%

\section{GaAs ultrathin solar cell design}
\label{sec:design}

\begin{figure}[htpb]
    \centering
    \includegraphics[width=1.0\linewidth]{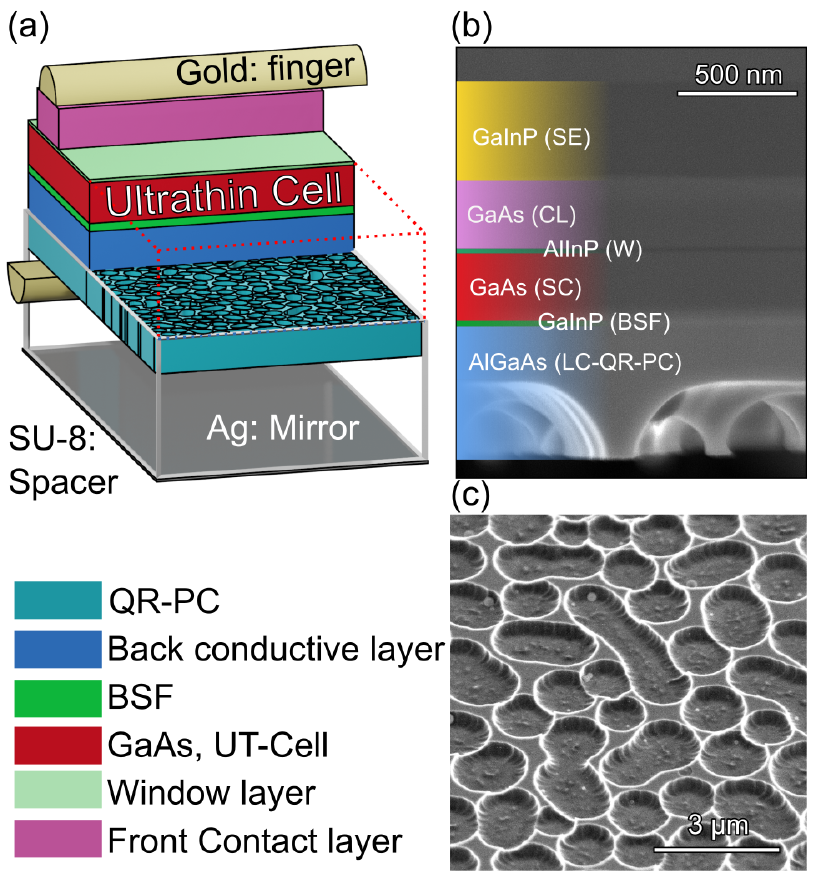}
    \caption{(a) Isometric cross sectional view of the fabricated ultrathin GaAs cell. The image shows a detail of the structure, not the full cell. (b) SEM cross sectional view of the cell after transferring the pattern to the semiconductor, prior to inversion. Each layer of the solar cells is false-color following (a). (c) Tilt-view (45 degrees) of a test quasi-random structure transferred to AlGaAs.}%
    \label{fig:scheme_etch}
\end{figure}

The cell design is based on a rear-heterojunction single-junction GaAs cell with an $n$ on $p$ configuration, with the addition of a light-trapping structure on the back side. The backside photonic crystal is more forgiving than frontside designs as the cell can then use a standard anti-reflective coating\cite{buencuerpo_2020_oe}. The layer structure, shown in Fig.~\ref{fig:scheme_etch}(a,b), is as follows: We use a front contact GaAs layer (CL) that is later removed by selective wet-etching, followed by a bilayer window of 15 nm of AlInP and 5 nm of GaInP. The absorber layer is only 260 nm of n-type GaAs, about an order of magnitude thinner than in a conventional, optically thick cell. The pn-junction is formed at the heterojunction with a p-type GaInP layer (a so-called back surface field layer, or BSF), thinned to 40 nm. Finally, we grow a relatively thick 540 nm layer of Al$_{0.7}$Ga$_{0.3}$As in which we will pattern the QR structures, and a 20 nm back contact layer of Al$_{0.3}$Ga$_{0.7}$As. The thick AlGaAs has a high index of refraction that is necessary for a light-trapping structure\cite{buencuerpo_2020_oe}, and also serves as a lateral conduction layer to spread the current. Following QR patterning, as described below, we add an SU-8 dielectric spacer and a silver mirror to create the final the optical cavity, as we proposed for the computational studies\cite{buencuerpo_2020_oe, buencuerpo_2021_olt}. The SU-8 dielectric spacer serves mainly to mitigate parasitic absorption of the out-of-normal light diffracted by the quasi-random photonic crystal, but has the secondary advantage of planarizing the patterned semiconductor surface before it is covered with the silver, thus lowering the roughness on the mirror.

The cell was designed to maximize the transparency of all of the semiconductor layers surrounding the GaAs absorber. The transmission after passing through the 260 nm GaAs absorber will define the spectrum for the rear layers. The upper layers (ARC, window) are identical to a thick GaAs device. However, the layers in the back are the most critical to achieving a higher \Jsc while keeping the electrical design of the cell intact. The thin GaInP BSF and the high Al content in the back contact layer are the two main differences compared to a thick cell. The GaInP will parasitically absorb part of the incoming light after the single pass through the 260 nm layer. The Al$_{0.7}$Ga$_{0.3}$As layer will be almost transparent for the useful wavelengths longer than 550 nm. The thin layer with lower aluminum content is a compromise to obtain a lower contact resistance, even if it will absorb part of the light. We use AlGaAs instead of, for example, dielectrics, because of its high refractive index, $n\approx3.5$ compared to the surrounding spacer, $n\approx1.5$. A high contrast in $n$ will increase the diffraction efficiency of the structures and, consequently, its light trapping performance. However, the light-trapping effect (by increasing the optical path) of these structures will maximize the absorption for all the layers on the cell, without distinguishing which part of the cells is the active region. Therefore, at the operating wavelengths the absorption coefficient, $\alpha$, has to be close to zero for all layers in the cell besides the main absorber layer. This condition is even more critical for the PC material, as spatial light-confinement is also expected, otherwise it will parasitically absorb light\cite{buencuerpo_2020_oe}. %
 
\begin{figure*}[htpb]
    \centering
    \includegraphics[scale=0.8]{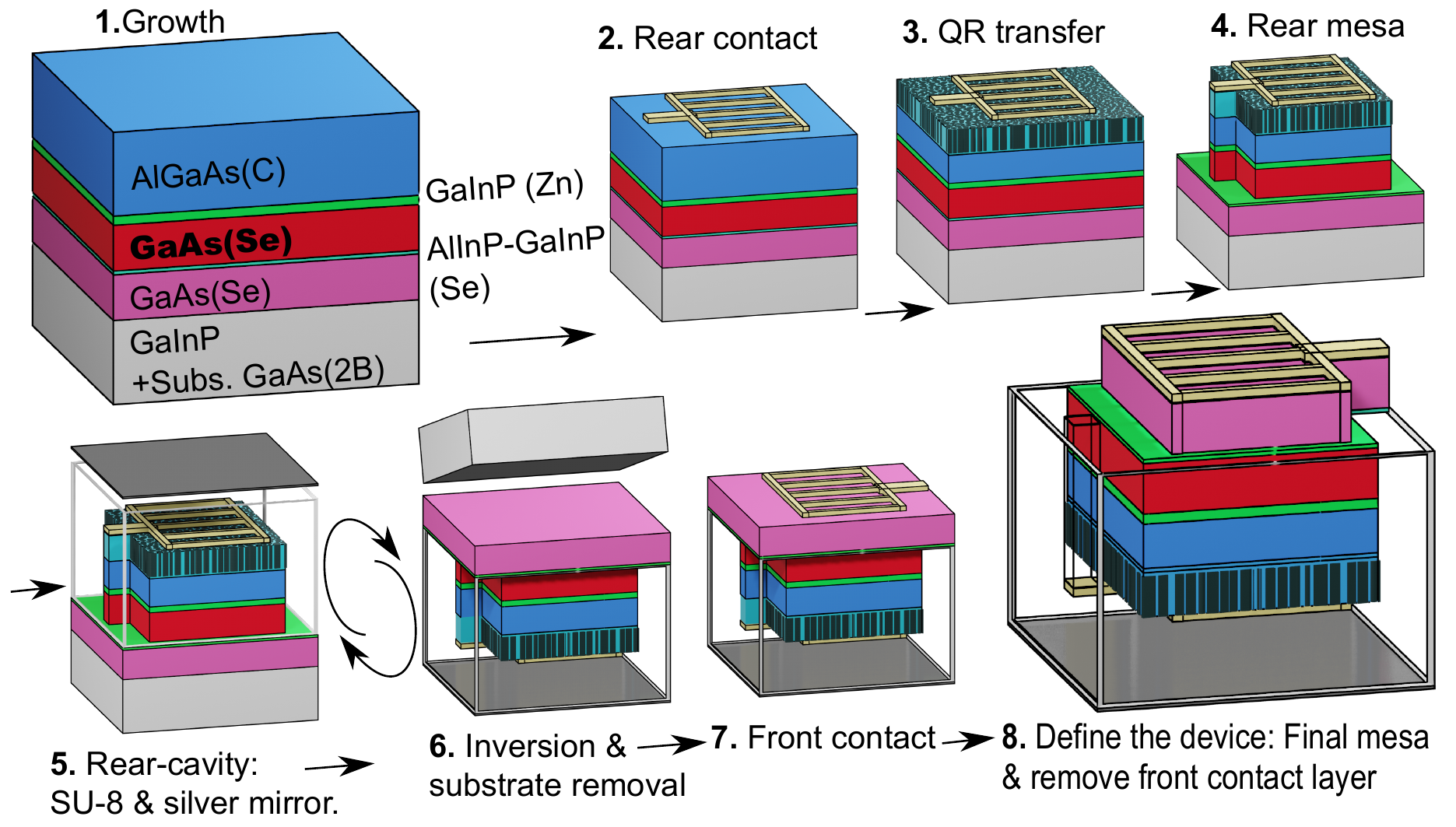}
    \caption{Scheme of the fabrication process, (1) the cell is grown using MOCVD, (2) The rear-contact is created by lithography and electroplated, (3) The QR process, as in Fig. \ref{fig:qr_method} is used to create and transfer the pattern, (4) Define the device until the window layer, (5) spin-coating SU-8 and silver to create the rear-cavity, (6) inversion of the sample by gluing the sample to a glass holder and substrate removal, (7) The front electroplated fingers are created following the same method as in the rear fingers, (8) removal of the extra semiconductor around and on top of the bottom pad of the device. The front contact layer is removed using the front grid fingers as a mask.}%
   \label{fig:inverted_process}
\end{figure*}

After growth, the semiconductor is processed into individual devices following a variation of the standard inverted process\cite{duda_2012_}, shown in Fig.\ref{fig:inverted_process}. Briefly, the gridded back gold contact is electroplated to the back contact layer (2); the QR structure is patterned on the back (3), as will be described in detail in the next section; the device is mesa-etched from the back side (4) down to the window layer, and the SU-8 and 350-nm silver mirror are deposited (5). Then the structure is bonded with epoxy to a silicon handle and the substrate is etched away (6); the front grid is electroplated (7), in alignment with the back grid; the remainder of the mesa is isolated and the front contact layer is removed (8). In some devices, a bilayer MgF$_2$/ZnS (105/50 nm) iss deposited at the end by thermal evaporation. More details can be found in the Appendices.

\section{Fabrication Methods: Polymer blend for maskless fabrication of quasi-random photonic crystals}
\label{sec:fab_qr}
We use poly-methyl-methacrylate (PMMA) and polystyrene (PS) with similar molecular weights (M$_w\approx$ 100 Kmol/g) to form the polymer blend. Then, we use propylene glycol monomethyl ether acetate (PGMEA) as a solvent for both polymers. PGMEA is typically used in conventional photoresists and standard UV lithography processes. The common solvent is evaporated by spin-coating and spinodal decomposition occurs\cite{swalheim_1999_s}. 

\begin{figure*}[htpb]
    \centering
    \includegraphics[width=0.8\linewidth]{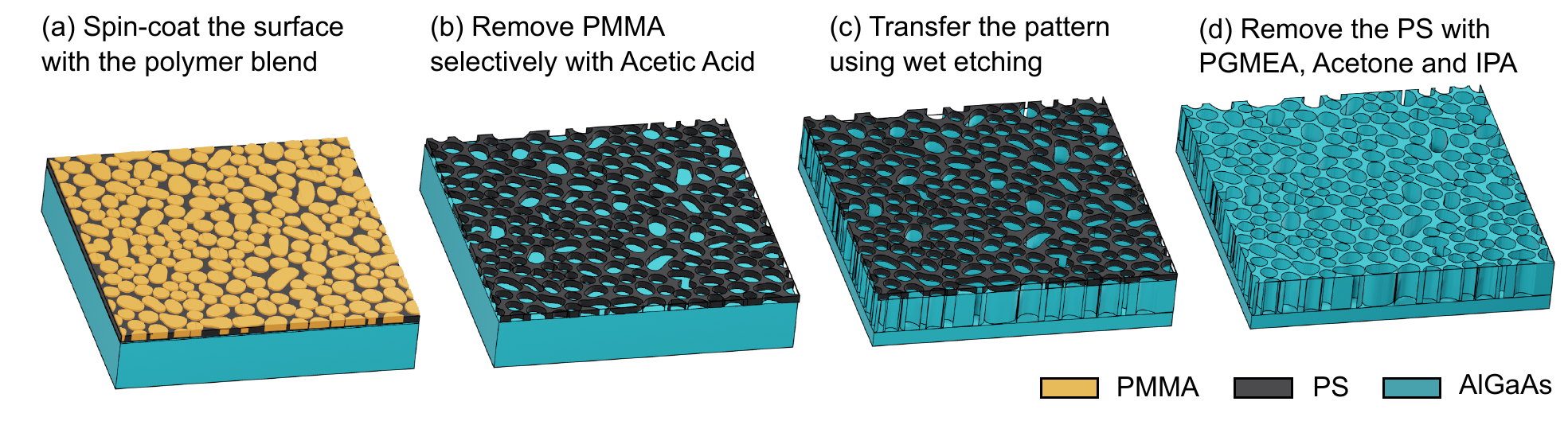}
    \caption{Process used for creating the QR pattern. (a) Surface after the spin coating solution with islands of PMMA (b) Surface after removing the PMMA using acetic acid, (d) the pattern transfer to the AlGaAs using wet etching, (d) removal of the PS with PGMEA, acetone and IPA.}%
    \label{fig:qr_method}
\end{figure*}

Once the solvent evaporation freezes the pattern, selective wet etching removes only one of the polymers, leaving the other as a pattern to be transferred into the semiconductor. 
We use selective wet etching instead of reactive ion etching (RIE) to transfer the polymer pattern into the semiconductor. The process leaves the PS unetched, using cheaper wet chemistry, and avoids ion damage in the solar cell. However, as we use an isotropic etch, the 2D pattern maximum thickness is half the average lattice. Furthermore, the borders of the structure present an apodized base because of the under etch, shown in Fig.\ref{fig:scheme_etch}(c). The success of the wet etching depends on having enough adhesion between the polymer and the semiconductor to survive the wet-etching process, and obtaining lateral polymer-oriented domains. If the domains of the PS are not vertically oriented, the wet etching will not reach the III-V material, and, therefore, the pattern will not be transferred. After successfully etching, we remove the PS using consecutive baths in PGMEA, acetone, and IPA, leaving behind the pattern in the AlGaAs layer.
With this reproducible polymer blend lithography process, we can explore how different structures are created by modifying the polymer blend coating conditions.

\section{Results}
\subsection{From a random microstructure to quasi-random photonic crystal}
Controlling the size of the structures and their filling fraction is fundamental to achieving high scattering in the desired wavelengths, and therefore higher absorption in the ultrathin GaAs cell. The reciprocal space can be used to identify the best structures for light-trapping. Once the polymers and solvent are set, the process is controlled by the thickness of the coating and the etching time to define the thickness of the structure. The thickness of the coating depends on the dilution concentration and the spin-coating speed, $s$. $s$ also defines the in-plane average lattice of the pattern. However, it is important to acknowledge that the thickness of the polymer coating is irrelevant to the later transferred nanostructured \emph{thickness} when using a perfectly selective etching, as we have between the PS and the III-V. These control knobs are used to create a 3D structure on a planar surface of the semiconductor. The overall process is depicted in Fig.~\ref{fig:qr_method}.

The ratio between PMMA and PS will shift the real space from PS domains surrounded by PMMA (for ratios of PS lower than PMMA) to PMMA domains surrounded by PS\cite{swalheim_1999_s}, following the Cahn-Hilliard equation. We initially use a 1:1 ratio in weight between polymers\cite{buencuerpo_2021_2i4pscp}.
Still, after etching, due to the isotropic nature of the wet etching we use in our process, the under-etch lowers the photonic crystal's overall filling fraction. Intuitively, if the pattern is very shallow or with a low density of scatterers, those structures are not good candidates for light trapping. Therefore, to compensate for the under-etch, we use an 11:9 ratio (PS:PMMA) to obtain slightly larger PS domains. 

\begin{figure*}
    \centering
    \includegraphics[width=\textwidth]{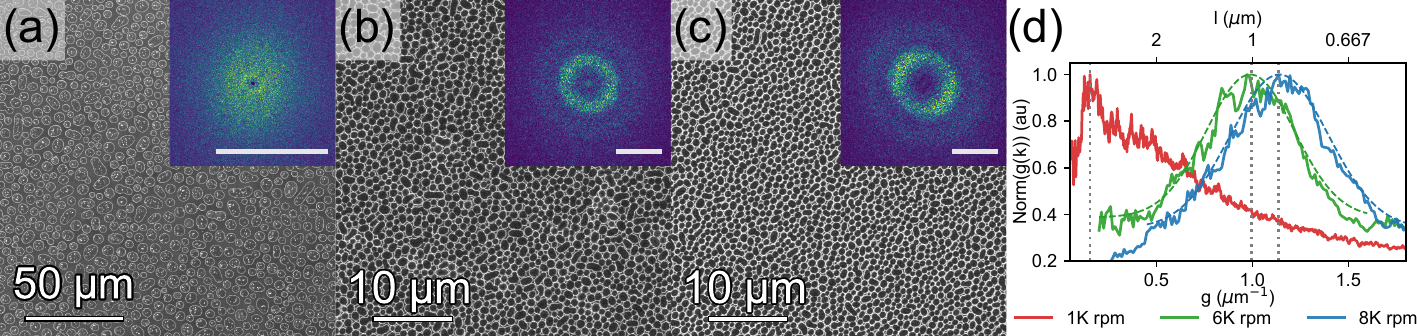}
    \caption{SEM images of the quasi-random patterns transferred to the AlGaAs conductive layer in the back of the solar cell; inset: FFT of image to define the average lattice parameter, scale bar 2$\mu$m$^{-1}$. The PB is spin-coated at 1K rpm (a), 6K rpm, (b), 8K rpm (c). (d) The radial profile of the reciprocal space, $g(k)$, normalized to the peak maximum for: $s$=1000 rpm a (red), and $s$= 6000 rpm (green), and $s$=8000 rpm (blue).} \label{fig:qr_k}
\end{figure*}

The thickness of the PB coating is thinner with higher spin-coating speeds, $s$, as occurs with all photo-resists. The higher the speed, the thinner the coating, and consequently the smaller the polymer blend domains\cite{zhang_2018_oeo}. We use different spin-coating speeds to modify the lattice and pattern of the structures. We explore three spin-coating speeds: a low speed at 1000 rpm and two high-speed spin-coatings, 6000~rpm and 8000~rpm. The patterns obtained are shown in Fig.\ref{fig:qr_k}. We study the ordering by analyzing the reciprocal space and its spatial frequencies to determine which structures are QR-PCs, using the Fourier transform (2D) and the average radial reciprocal space, $g(k)$, (1D). Then, we perform a fitting to a normal distribution around the ring's peak, as shown in Fig.\ref{fig:qr_k}(d) with fitting parameters in Table \ref{tab:morphology}.
\begin{table}
\centering
\caption{Quasi-random feature sizes: Center spatial frequency, $g_{\mu}$, obtained by analyzing the PSD, $g(k)$, by finding its maximum for 1K rpm or fitting the radial PSD, $g(k)$, to a Gaussian, with half-width, $\sigma$ for the rest of cases; the average lattice $L$=1/$g_{\mu}$.}
\begin{tabular}{lccr}
$s$ (rpm) &  $g_{\mu}$ ($\mu$m$^{-1}$)  &  $\sigma$ ($\mu$m$^{-1}$) &  L ($\mu$)m \\
\hline
1000  & 0.19 & - & $\approx$ 5.26 \\
6000  & 0.99 &   0.32 &  1.01 \\
8000  & 1.14 &   0.38 &  0.88 \\
\hline
\end{tabular}
\label{tab:morphology}
\end{table}

The structures transferred to the AlGaAs, shown in Fig.\ref{fig:qr_k}, exhibit different average sizes, with bigger structures close to 5-10 $\mu$m for the 1000 rpm case, and smaller size patterns for the 6000 and 8000 rpm patterns. When analyzing the reciprocal space, we see two spatial frequency distributions. We do not observe the typically defined ring of hyper-uniform structures for the lower speed, as shown in the inset of Fig.\ref{fig:qr_k}(a). In contrast, the two higher speeds of 6000 rpm and 8000 rpm present a clear ring around the zero-order in the reciprocal space, as we show in Fig.\ref{fig:qr_k}(b),(c). When integrating to obtain $g$ we can see the lower speed presents a maximum at the $g$=0.19 $\mu$m$^{-1}$, with an equivalent lattice of 5.26 $\mu$m. However, it presents a slow decay for higher frequencies and therefore smaller lattice parameters. The lower case presents micro-structures of a few micron with a lower packing fraction, and therefore, the Fourier space is randomly distributed without a clear localized PSD. When inspecting the 6000 rpm and the 8000 rpm $g$ we can see a different function. The $g$ can be fitted using a Gaussian profile. Using this fitting, we obtain the $g_\mu=1.01 \mu$m and $g_\mu=0.88 \mu$m for the 6000 rpm and 8000 rpm structures, respectively. Both structure present smaller ovaloids (or connected channels) with a defined average lattice with a clearly defined PSD. 

Micro-structures like the one created at 1000 rpm have the benefit of creating a broadband response due to the rough and non-localized reciprocal space. These patterns are expected to smooth interference patterns such as Fabry-Perot, thanks to the added resonances. Still, the expected intensity of these resonances is lower as there is no localized region in the reciprocal space. The scatterers are re-distributing the momentum across a continuum of orders. Structures with a localized PSD (as a quasi-random photonic crystal) and smaller features closer to the GaAs bandgap are more desirable for increasing the absorption in the ultrathin solar cell. Therefore the QR-PCs obtained at 6000 rpm and 8000 rpm are better candidates as light trapping structures for ultrathin GaAs cells.

\subsection{Ultrathin GaAs cells with rear quasirandom photonic crystals}

\begin{figure*}[htpb]
    \centering
    \includegraphics[]{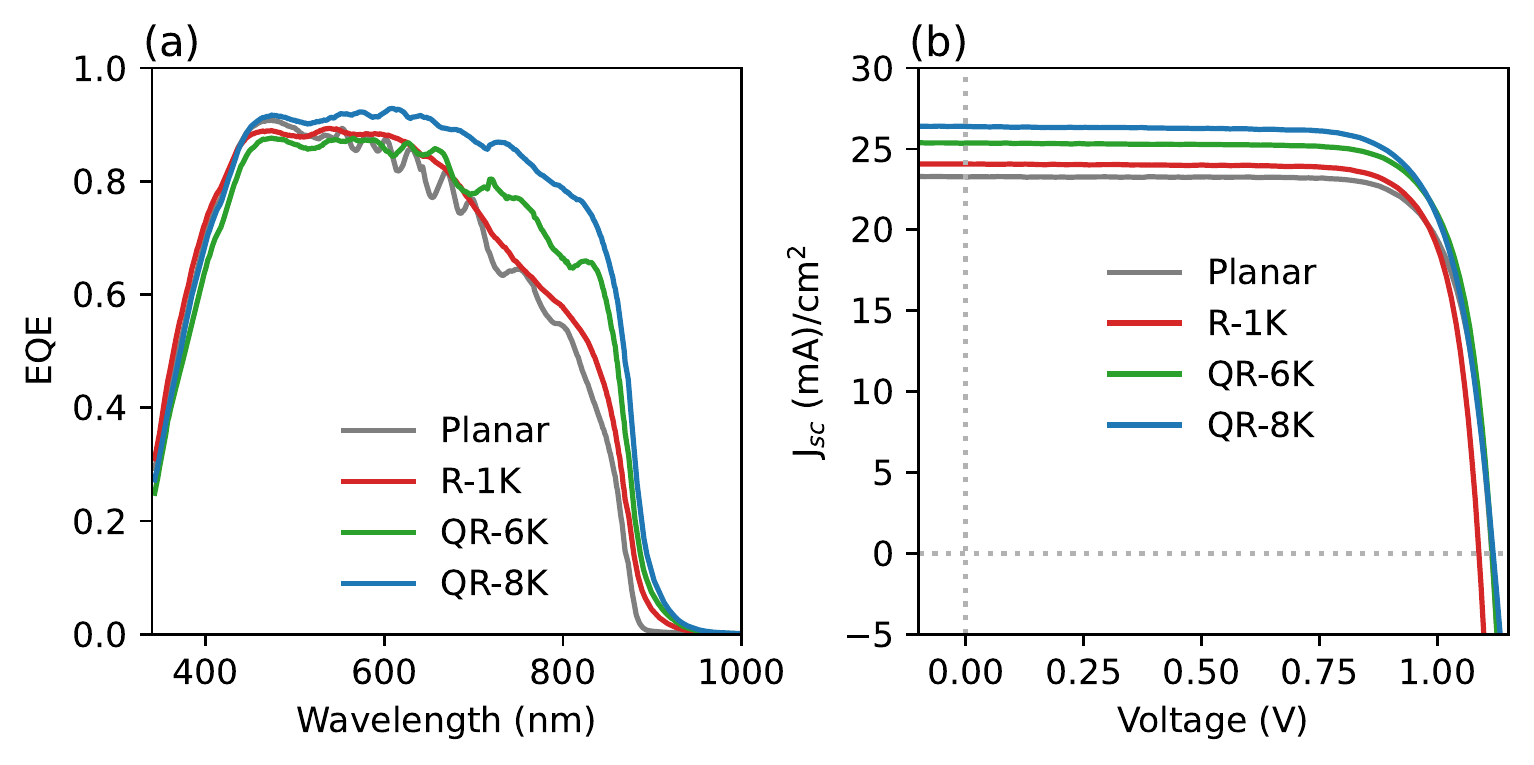}
    \caption{External quantum efficiency (EQE) (a) and current density vs voltage (b) for cells without patterns, i.e. planar (black), and with rear patterning created at 1000 rpm, R-1K, (red), and quasi-random photonic crystals created at 6000 rpm, QR-6K (green) and at 8000 rpm, QR-8K (blue).} %
    \label{fig:qe_iv}
\end{figure*}

Using the process described above, %
we implement four test cases processed in parallel, all with 260 nm absorber layers: a reference planar mirror case, a micro-structured random structure (R-1K, $s=$1000~rpm) and two quasirandom structures fabricated with $s=$6000~rpm (QR-6K), and with $s$=8000~rpm (QR-8K). 
The best-nanostructured devices with their reference cases are shown in Fig.\ref{fig:qe_iv}; their efficiencies and voltages are summarized in Table \ref{tab:qe_iv}.%
\begin{table*}[htbp]
  \centering
  \caption{Spin speed and electrical properties: \Jsc, \Voc , FF and $\eta$ for the four cases studied (all uncertainties in the supplementary information).}
    \begin{tabular}{lccccc}
    Case & $s$ (rpm) & \multicolumn{1}{l}{\Jsc (mA/cm$^2$)} & \multicolumn{1}{l}{\Voc} & \multicolumn{1}{l}{FF} & \multicolumn{1}{l}{$\eta$ (\%)}  \\
    \hline
    Planar&  -      & 23.3 & 1.115 & 78.3 & 20.33 $\pm$ 0.14 \\
    R-1K  &   1000  & 24.1 & 1.088 & 79.0 & 20.69 $\pm$ 0.14 \\
    QR-6K &   6000  & 25.4 & 1.116 & 77.6 & 21.97 $\pm$0.14  \\
    QR-8K &   8000  & 26.4 & 1.119 & 75.7 & \textbf{22.35} $\pm$0.15  \\
    \hline
    \end{tabular}%
  \label{tab:qe_iv}%
\end{table*}%

The planar cell of 260 nm presents the lowest $J_\mathrm{sc}$=23.28 mA/cm$^2$, and efficiency 20.33\%. The lower \Jsc is expected for this device, as no QR-PC modifies the light momentum. The only aid to increase the absorption is the rear spacer and silver mirror, which doubles the optical path followed by the incident light. The cell presents Fabry-Perot (FP) resonances that are minimized as we use an ARC, and can be understood as a very low-quality-factor photonic cavity. The R-1K case presents a small but nonetheless significant improvement in the QE compared to the planar cell, and a correspondingly higher $J_\mathrm{sc}$. The random structure averages the FP resonances, with a small improvement in the overall absorption.

The maximums of the FP resonances for the planar case at lower wavelengths are very close to QE from the random structures, R-1K, case. The improvement is not clear. We associate this to a non-localized reciprocal space with an overall increased absorption, but with less intensity. 
The QR-6K clearly surpasses this limit, with an overall improvement for all the regions with incomplete absorption. The efficiency is improved to 21.97\%, thanks to this increase in the cell absorption. Finally, the QR-8K presents the highest QE with an improvement of the \Jsc of 3.1 mA/cm$^2$ when compared to the planar device. The improvement in the QE is clear over the 550 nm to 900 nm range where the planar ultrathin-cell QE decays, and the efficiency increases to 22.35\%.
The \Voc does not vary between the planar, QR-6K and QR-8K, which came from the same epitaxial sample. The R-1K sample was grown earlier and has a lower \Voc but higher FF. Using the same epitaxial sample as R-1K, an equivalent planar device for reference and a QR-6K were fabricated and have a comparable QE to the best cases presented in Fig. \ref{fig:qe_iv}, along with the same \Voc and FF as the R-1K sample. This data are shown in the supplementary information.

\section{Discussion}
When inspecting all of the devices, there is a clear trend of increasing $J_\mathrm{sc}$ when increasing the spin-coat speed, leading to a more localized reciprocal space ring with smaller features, as shown in Fig.\ref{fig:qe_iv}. The QR-PCs (QR-6K, QR-8K) clearly outperform the planar case and the random micro-structure (R-1K) in \Jsc. The relative increase in the \Jsc is 13\%, whereas variations for the FF are below 5\% and variations in \Voc are insignificant. Thus the increase in the \Jsc is the main difference between samples, and is the main driver for the increasing efficiency with nanostructuring.

\begin{figure}[htpb]
    \centering
    \includegraphics[scale=1]{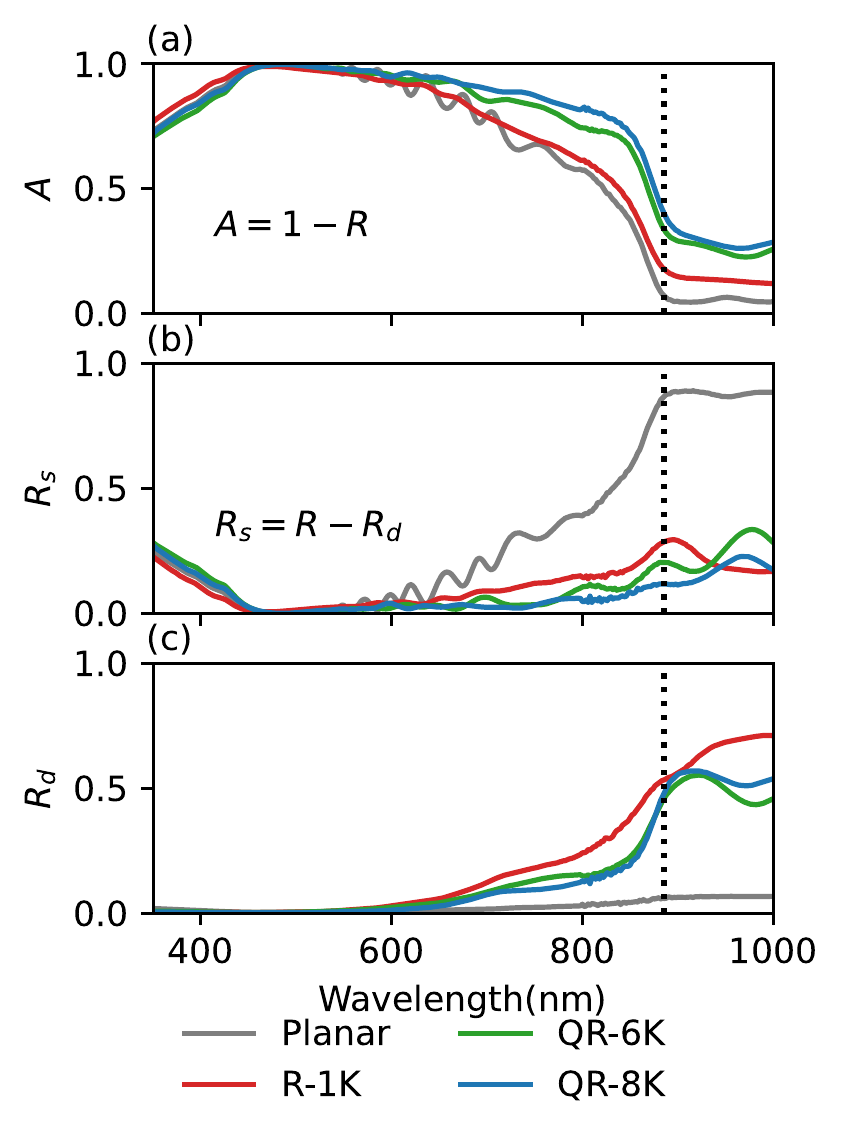}
    \caption{Reflectance measurements using an integrating sphere for the planar case (gray), R-1K (red), QR-6K (green), and the QR-8K rpm (blue): (a) the optical absorption, as $A=1-R$ for the sample using an integrating sphere for the four cases studied; (b) The specular component obtained by subtracting the scattered reflectance, $R_d$ from the total reflectance, $R$, measured when opening the port of the specular reflectance; (c) the direct measurement of $R_d$. GaAs bandgap wavelength dividing the spectrum (vertical dotted black).}%
    \label{fig:reflectance}
\end{figure}

The increase of \Jsc comes from the increase QE in the long-wavelength region, as shown in Fig.\ref{fig:qe_iv}(a), as was the goal of the design. This effect is purely optical. Physically, the QR-8K case is absorbing more light in the long wavelength region than the other cases. To demonstrate this, we have characterized the devices using an integrating sphere, with and without the specular exclusion port, thereby measuring both the total reflection, $R$, and the non-specular reflection (diffuse or diffracted due to the spectral frequencies), $R_d$. The specular component, $R_s = R - R_d$, is the only reflection component present in planar cells, but both components are present in diffractive structures. It is expected that the random microstructures and QR-PCs present a high scattered component, i.e., light being reflected out of the specular regime for wavelengths reaching the rear of the cell. Mathematically, this superposition of reflection components is analogous to the numerical analysis splitting the zero order, $g=0$, and the diffracted orders, $g>0$, with $R=R_{g=0} + R_{g>0}$ assuming diffracted angles are bigger than the open port on the integrating sphere \cite{buencuerpo_2020_oe}. Also, as we have an optically thick metallic mirror, we can assume the optical absorption of the system is $A=1-R$. The results are presented in Fig.\ref{fig:reflectance}.

For wavelengths near the bandgap, we observe that the optical absorption, $A=1-R$, is higher for the QR-PCs than for the other two cases, as we show in Fig.\ref{fig:reflectance}(a). The FP resonances on the planar case match the QE for its case. The R-1K averages the FP maximums, but it does not clearly surpass the planar case. For the QR-PCs, the higher absorption explains the higher QE for QR-6K and QR-8K. In the planar case, the main loss is $R_s$ whereas for the R-1K and the QR-PCs we have both $R_d$ and $R_s$ lowering the absorption of the cell. The increased $R_{g>0}$, and therefore $R_d$, is expected for QR-PCs \cite{buencuerpo_2021_olt}.

We proceed to analyze why the QR-PCs outperform the random microstructure. 
The reflection loss for diffracted light, $R_{d}$, arises from light that has exchanged momentum with the structures (after one or more passes through them), but at an angle is not high enough to be trapped inside the cell structure. The R-1K structure can scatter a high percentage of the incident light, with high $R_d$. However, this light has an angle lower than the escape cone, and therefore is not trapped: $R_d$ is light not trapped in the cell. Therefore, analyzing the haze or $R_d$ cannot predict by itself the light-trapping capabilities of a structure. Using $R_s$ we observe that the specular component is higher for the R-1K than the QR-PCs for wavelengths near the bandgap. The QR-PCs present a lower $R_s$ in the incomplete absorption region than the R-1K because they diffract light more efficiently. However, this light does not escape the cell through $R_d$ (the light has a higher angle than the escape cone). The combination of higher diffraction efficiency, lowering $R_s$, and at a higher angle that lowers $R_d$, has the effect of increasing the absorption of the ultrathin cell for the QR-PCs. The increased absorption is translated in a higher \Jsc and a higher efficiency.

The cells presented in this work are more efficient than recent ultrathin cells in the literature, with 22.35\% compared to 19.9\% by Chen et al. \cite{chen_2019_ne}, and 21.40\% by Eerden et al. \cite{eerden_2020_ppra}. Those structures are reported with 205 nm and 300 nm absorbers, respectively. Beyond efficiency, each work presents different approaches for the rear structure of an ultrathin GaAs cell: Chen et al. presented a simple-periodic (plasmonic) structure, whereas, Eerden et al. presented rough surfaces created by wet etching without ordering. This work demonstrates quasi-random photonic crystals fabricated with polymer-blends and wet-etching, which lies in between both approaches and targets the most critical spectral range. Also, the QR-PCs can be optimized for other cells by targeting smaller or larger average lattices like a simple photonic crystal in contrast to purely anisotropic wet-etching, but it does not require masks for each lattice. Finally, the work presented here surpasses the \Jsc of both works by approximately 2 mA/cm$^2$, showing the efficient light-trapping properties of QR-PCs created by this method.

\section{Conclusion}
In this work, we demonstrate a fabrication method based on polymer blends that can be used to create light-trapping structures in ultrathin GaAs cells. %
We use this process in conjunction with wet etching to transfer the pattern to the semiconductor, to create a photonic crystal. The structure of the patterns is controlled by spin coating. We create quasi-random photonic crystals with a localized Gaussian ring in the reciprocal space, and reach average lattice parameters as low as 880 nm. We apply them to an ultrathin GaAs cell, with an absorber thickness of 260 nm. The QR-PC obtains a high QE over all the absorption region, with a \Jsc comparable to thick devices. We obtain a certified photocurrent of 26.38 mA/cm$^2$ (3 mA/cm$^2$ more than the planar device), 89\% of the current of a record thick device (29.6 mA/cm$^2$). The best device reaches 22.35\% efficiency (2\%abs more than the planar reference). The rear photonic crystal created by this process does not damage the final \Voc of the cell, with a high 1.12 V directly comparable with thick GaAs cells with excellent reflectors. Future works centered on improving the FF of these devices to values closer to a 85\%, either by improving the lateral resistance of the rear-conduction layer and/or adding more back grid fingers will move these devices to the 25\% efficiency range. The work presented here enables the use of complex photonic crystals on III-V materials by using spin-coating and wet etching. 

\appendix
\section{Device growth and processing}
\label{sec:fab_cell}
The cell is grown by atmospheric pressure Metal-Organic Vapor Phase Epitaxy (MOVPE) in a custom-built reactor. The cells are grown on silicon-doped (001) GaAs substrates, miscut 2 degrees toward the (111)B direction. The dopants used for each layer are detailed in Fig.\ref{fig:inverted_process}(a).
The general process is described in Fig.\ref{fig:inverted_process}. The process is a modification of the inverted process described in Ref.\cite{duda_2012_}. The first step after growth is to generate the rear-fingers. We spin-coat the back of the cell and cover the back of the substrate with photoresist (S1818 at 4000 rpm). Optical lithography defines the fingers. The metal is deposited by electroplating gold. The quasi-random pattern is transferred following the details in Section \ref{sec:fab_qr}. Once the pattern is transferred and PS is removed from the surface, the fabrication process can continue. We do a rear-mesa etch until the window layer, by consecutive selective wet etchings using concentrated HCl for phosphide layers and H$_3$PO$_4$ : H$_2$O$_2$ : H$_2$O (3:4:1 by volume) for arsenide layers \cite{duda_2012_}. Then we spin-coat SU-8 at 3000 rpm for 30 s, and soft-bake at 95 C. A flood exposure under UV (15~s, 35~mJ/s), post-bake (90~s), and development step (60~s) are done to ensure the cross-linking of the photo-resist. Then the sample is dehydrated and hard-baked in the oven at 100 C overnight. We evaporate 350 nm of Ag in the rear of the cell using an electron-beam evaporator. Then, we proceed to do a standard inversion, using epoxy (Tribond-931) to bond the cell's rear side (the silver) to a glass substrate. The GaAs substrate is removed by wet etching in NH$_4$OH : H$_2$O$_2$ (1:3), stopping selectively on the GaInP stop-etch layer (SE). The front fingers lithography is aligned with the back fingers as they are visible through the thin contact and window layer. The process is equivalent to the back-fingers done previously. Finally, the cell area is defined by doing a front mesa isolation etch and removing the extra semiconductor on top of the back contact. The front contact layer is selectively removed using a diluted mixture of NH$_4$OH : H$_2$O$_2$ : H$_2$O (2:1:10), selectively over the window layer and using the metal fingers as a mask.

\subsection{Polymers' blend fabrication details}
\label{sec:fab_etching}

We use 99.5\% concentration PGMEA as our solvent. We have found that PGMEA produces smoother and more homogeneous films than using tetrahydrofuran (THF) \cite{zhang_2018_oeo}. PGMEA has a lower vapor pressure than THF or methyl ethyl ketone, which may be the reason for the better homogeneity we observed. However, it is a worse solvent for PMMA\cite{gaikwad_2016_oe}, so higher concentrations are difficult to obtain. We fix the concentration at 0.25 mg/ ml in this study. 

The pre-conditioning of the surface is done using NH$_4$OH 30\%, and baking the substrate at 100 C on a hotplate for 5 min, or by immersion in glacial acetic acid for 30~s. Both approaches have proven successful for spin-coating speeds equal to or under 6000 rpm, however above 6000 rpm, only the pre-conditioning with acetic acid has worked. %

The samples were spin-coated at different speeds. Once the polymer pattern is defined, we proceed to do a softbake at 130 C for 10 min. Then, we selectively etch for 3 min in glacial acetic acid to remove the PMMA and leave the PS pattern. Before transferring the pattern we do an additional hardbake of 5 min at the same temperature.

To transfer the structure to the rear AlGaAs layer in the cell, we etch the samples in diluted H$_3$PO$_4$ and H$_2$O$_2$ in DI water (3:4:10) (an etch rate of 35-40 nm/s is calibrated over 10 $\mu$m step defined by lithography). We kept the etching time constant at 6-7~s, controlled manually. %
The PS is removed after the wet etching by consecutive etches in PGMEA, IPA, Acetone, and IPA. %

\section{Cells characterization}
The cells studied in the main text were independently certified by the NREL Cell and Module Performance (CMP) team.  The areas of the devices were clearly defined to be about 0.25 $cm^2$ by the mesa etching described in the processing section. The actual area of each device, reported in the certified results that are compiled in the supplementary information (SI), was measured under a dedicated microscope system with calibrations governed by ISO certification using standard protocols for area measurement. The total mesa area was used, including all front metal. No separate aperture was used as the area was completely defined by the mesa isolation.
 
The external quantum efficiency (EQE) was measured with chopped monochromatic light from a three-grating monochromator covering a range of 280 - 2000 nm, with a DC light bias of approximately 1/3 sun as detailed in the SI. Current-voltage (IV) curves were measured under an X-25 solar simulator with a Xe lamp. The illumination was adjusted to the one-sun AM1.5 global spectrum (G173 standard) at 1000 $W/m^{2}$ using a primary calibrated GaAs reference cell and spectral mismatch correction. The samples were controlled to 25°C on a vacuum hold-down stage. IV curves were swept from 102\% of $V_{oc}$ forward bias to -20\% of $V_{oc}$ reverse bias in a fast measurement with hysteresis check. No hysteresis nor change in results over a period of several months were observed as is typical of III-V photovoltaics, though a stability analysis was not specifically performed. No preconditioning was performed, but the $V_{oc}$ was measured without load to determine the IV range. Only a single device of each condition was measured, so no statistical information was obtained, but uncertainties were determined from typical measurement resolutions and repeatabilities.

\section{Image Analysis}
We do the Fourier transform of the SEM image, using a Lanczos envelope to mitigate the border effects \cite{necas_2012_c}, shown in Fig.\ref{fig:qr_k} insets. We generate the reciprocal space (2D), and we generate the radial integrated reciprocal space, $g(k)$, to generate the radial distribution of the structure. 
\section{Reflection Measurements}
Total and diffuse reflectivity measurements are collected on a Cary7000 with external diffuse reflectance accessory DRA2500 integrating sphere. The 150 mm sphere operates at 8$^\circ$ angle of incidence with the sample at the rear exit port. The diffuse-only reflectivity is measured by removing the specular exclusion port plug, which excludes $\sim 15^\circ$ around the specular reflection. A small spot kit and 5 mm aperture is used to illuminate only the active area of the cell. Total reflectivity was calibrated with a NIST aluminum specular reflectivity standard.

\subsection*{Acknowledgments}
We thank Jeff Carapella and Waldo Olavarría for growing the samples used in this work. We also thank Bill McMahon and Ryan France for helpful discussions. We also thank Tao Song for certifying the the device performance.
We thank these crazy pandemic times for letting us re-think all our processes for several months while working from home, then forcing us to modify all our processes; repetition and patience are the mothers of science.
This work was supported by the U.S. Department of Energy under Contract No. DE-AC36-08GO28308 with Alliance for Sustainable Energy, LLC, the Manager and Operator of the National Renewable Energy Laboratory. Funding provided by U.S. Department of Energy, Office of Energy Efficiency and Renewable Energy, Solar Energy Technologies Office under Agreement Numbers 34358 and 34911. The U.S. Government retains and the publisher, by accepting the article for publication, acknowledges that the U.S. Government retains a nonexclusive, paid up, irrevocable, worldwide license to publish or reproduce the published form of this work, or allow others to do so, for U.S. Government purposes.
\subsection*{Competing interests}
The authors declare that they have no known competing financial
interests or personal relationships that could have appeared to influence
the work reported in this paper.


\end{document}


\title{Supplementary Materials: Efficient light-trapping in ultrathin GaAs solar cells using quasi-random photonic crystals}

\author[nrel,ipvf,c2n]{Jeronimo Buencuerpo\corref{cor1}}
\ead{jeronimo.buencuerpo@ipvf.fr}
\author[nrel]{Theresa E. Saenz} 
\author[nrel]{Mark Steger} 
\author[nrel]{Michelle Young}
\author[nrel]{Emily L. Warren}
\author[nrel]{John F. Geisz}
\author[nrel]{Myles A. Steiner}
\author[nrel]{Adele C. Tamboli}
\cortext[cor1]{Corresponding author}

\address[nrel]{National Renewable Energy Laboratory (NREL), Golden 80401 Colorado, USA}
\address[ipvf]{Present address at L'Institut Photovoltaïque d'Île-de-France (IPVF), Palaiseau, France}
\address[c2n]{Present address at Centre de Nanosciences and Nanotechnologies (C2N), CNRS, Paris-Saclay University, Palaiseau, France}


\maketitle

\section{Integrated QE photocurrent measurements}
~
\begin{table}[htpb]
\centering
\caption{Comparison between the integrated EQE $J_\mathrm{sc}$ from the certified measurements an the integrated EQE using AM1.5g spectrum normalized to 1000 W/m$^2$.}
\begin{tabular}{lcl}
Case   & Integrated EQE $J_\mathrm{sc}$ &  $J_\mathrm{sc}$   \\
\hline
Planar & 23.00  & 23.28 $\pm$ 0.15 \\
R-1K   & 23.96  & 24.07 $\pm$ 0.16 \\
QR-6K  & 24.96  & 25.36 $\pm$ 0.17 \\
QR-8K  & 27.29  & 26.38 $\pm$ 0.17
\end{tabular}
\end{table}

\section{Comparison between epitaxies}
The epitaxies used for the cells study are named WB808 and WB872. WB808 is the epitaxy used for the R-1K case, with a higher FF, and lower \Voc, where as WB872 was the one used in the main text for the planar, QR-6K, and the QR-8K. WB808 epitaxy was also processed  as a planar reference, and 6 Krpm  QR, with slightly lower efficiency. The figure shows the Comparison between growths (WB808 and WB872) measured using a in-house QE system, and a solar simulator based on a Xe-lamp simulating a AM1.5g. The results are shown in Table S1. The QE system has a spot with a approximate diameter of 2 mm.  The cells were measured back to back, to minimize variations on the calibration due to lamp fluctuations. 

\begin{figure}[htpb]
    \centering
    \includegraphics[width=0.9\linewidth]{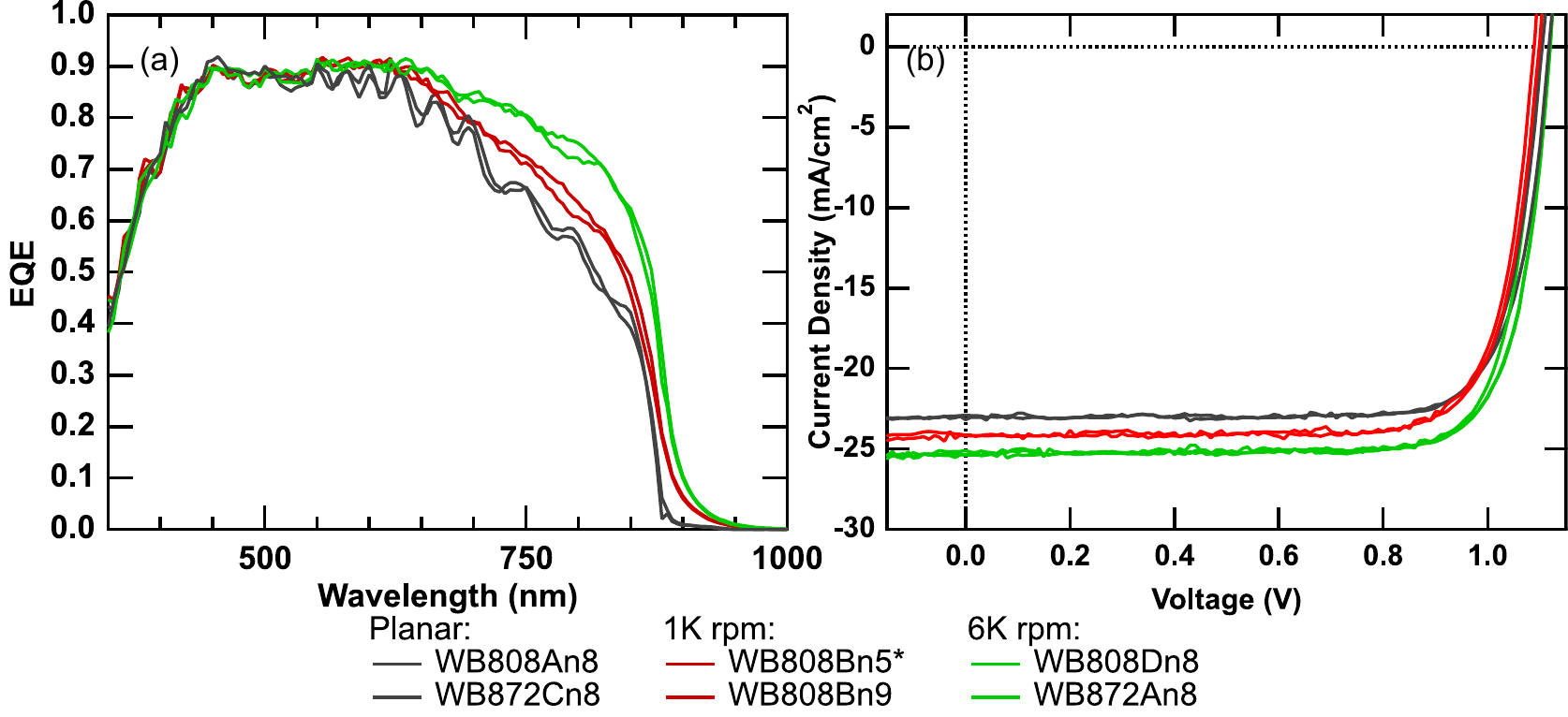}
    \caption{(a) External quantum efficiency and (b) JV curves for 6 devices comparing both epitaxies.}%
    \label{fig:all_qe}
\end{figure}

The Fabry-Perot oscillations for both epiaxies (WB808A and WB872C) shows an match between epitaxies. FP resonances depend on the thickness of the epitaxy, the processed SU-8, and the evaporated antireflective coating. A good match between both implies a reproducible thickness between samples, and therefore comparable \Jsc. The  6K rpm case, using both epitaxies (WB808Dn8 and WB872An8) shows a good agreement in the thickness and processing, more clearly on the \Jsc and IV measurements, see Table \ref{tab:iv_all}.

\begin{table}[htpb]
    \centering
    \caption{Summary of the electrical properties: \Jsc, \Voc , FF and $\eta$ comparing both epitaxies. Non-certified results.}
    \label{tab:iv_all}
\begin{tabular}{lccccccl}
Epitaxy & Cut & Dev. & Rpm      & Jsc (mA/cm2) & Voc   & FF    & \textbf{Eta} \\
\hline
WB872   & C   & n8   & -        & 23.19        & 1.117 & 78.81 & 20.42        \\
WB808   & A   & n8   & -        & 22.91        & 1.104 & 80.63 & 20.39        \\
WB808   & B   & n7   & 1000 rpm & 24.16        & 1.098 & 78.67 & 20.87        \\
WB808   & B   & n9   & 1000 rpm & 24.13        & 1.089 & 78.61 & 20.65        \\
WB808   & D   & n8   & 6000 rpm & 25.11        & 1.100 & 80.89 & 22.34        \\
WB872   & A   & n8   & 6000 rpm & 25.39        & 1.120 & 79.05 & 22.47       
\end{tabular}
\end{table}

The WB808 growth did have sligtly lower \Voc ($\approx$0.02V). As all the structures in this epitaxy, planar or with patterns, present this lower voltage, we associate this variation to a problem during growth. %

\section{Certified Measurements}
~
\newpage
\includepdf[ frame=true, scale=0.83,pages={2-9},]{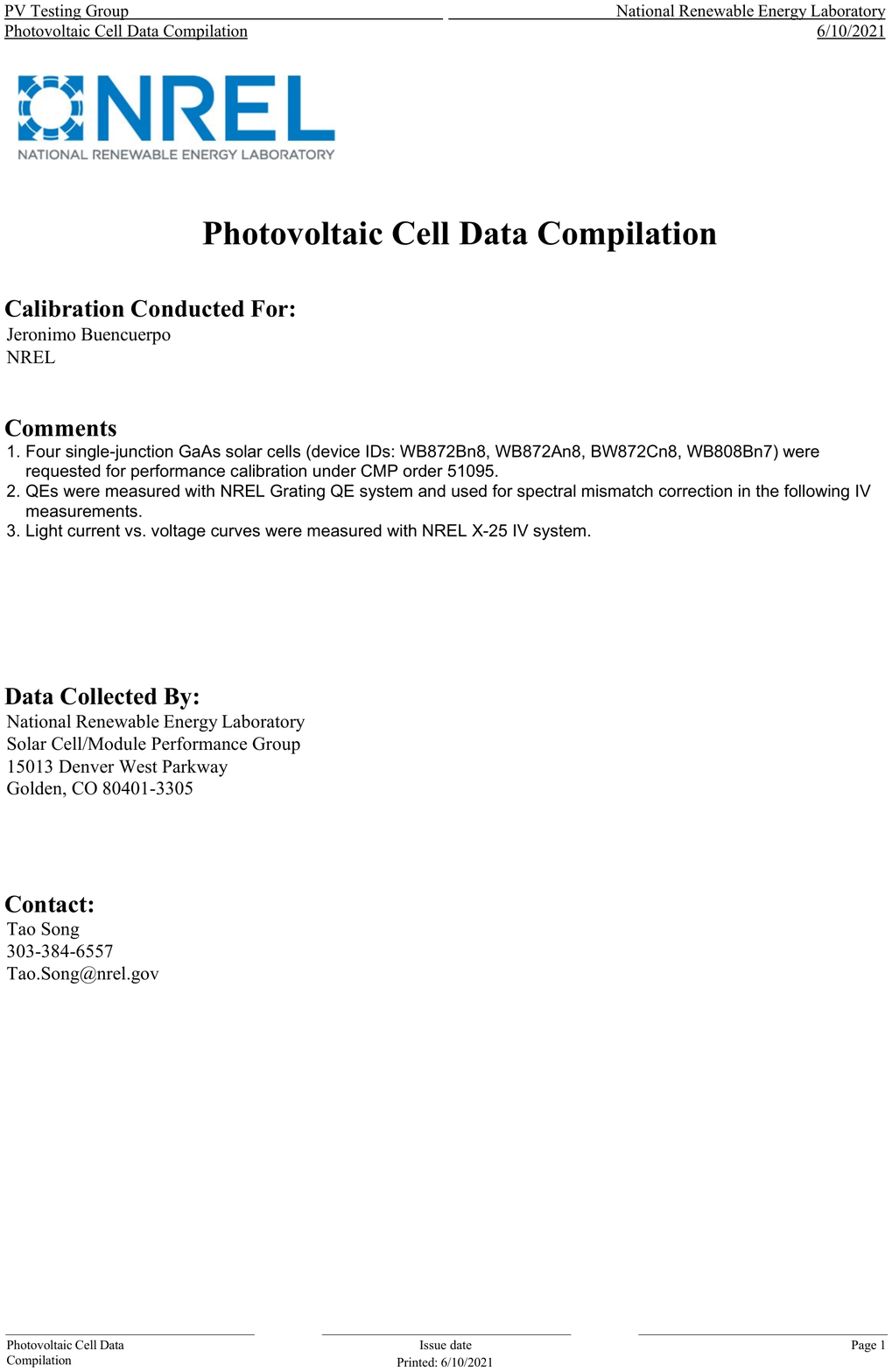}